# Enhanced Catalytic Activity of Gold@Polydopamine Nanoreactors with Multi-compartment Structure Under NIR Irradiation

Shilin Mei[1], Zdravko Kochovski[1], Rafael Roa[2], Sasa Gu[3], Xiaohui Xu[1], Hongtao Yu[1], Joachim Dzubiella[4,5], Matthias Ballauff[1,6], Yan Lu[1,7] ✉

✉ Yan Lu, yan.lu@helmholtz-berlin.de

[1] Soft Matter and Functional Materials, Helmholtz-Zentrum Berlin für Materialien und Energie, Hahn-Meitner Platz 1, 14109 Berlin, Germany

[2] Department of Applied Physics I, University of Málaga, 29071 Málaga, Spain

[3] College of Materials Science and Engineering, Nanjing Tech University, Nanjing 210000, People's Republic of China

[4] Institute of Physics, University of Freiburg, 79104 Freiburg, Germany

[5] Simulation of Energy Materials, Helmholtz-Zentrum Berlin für Materialien und Energie, Hahn-Meitner Platz 1, 14109 Berlin, Germany

[6] Institut für Physik, Humboldt-Universität zu Berlin, Newtonstr. 15, 12489 Berlin, Germany

[7] Institute of Chemistry, University of Potsdam, 14476 Potsdam, Germany

## HIGHLIGHTS

- Gold@polydopamine particles with multi-compartment structure were synthesized by using block copolymer of PS-*b*-P2VP as soft template and characterized by 3D electron tomography technique.

- The particles can be applied as catalytic nanoreactors for the full kinetic study of reduction reaction of 4-nitrophenol by NaBH$_4$.

- The nanoreactors show remarkable enhancement of the catalytic activity under NIR irradiation.

**ABSTRACT** Photothermal conversion (PTC) nanostructures have great potential for applications in many fields, and therefore, they have attracted tremendous attention. However, the construction of a PTC nanoreactor with multi-compartment structure to achieve the combination of unique chemical properties and structural feature is still challenging due to the synthetic difficulties. Herein, we designed and synthesized a catalytically active, PTC gold (Au)@polydopamine (PDA) nanoreactor driven by infrared irradiation using assembled PS-*b*-P2VP nanosphere as soft template. The particles exhibit multi-compartment structure which is revealed by 3D electron tomography characterization technique. They feature permeable shells with tunable shell thickness. Full kinetics for the reduction reaction of 4-nitrophenol has been investigated using these particles as nanoreactors and compared with other reported systems. Notably, a remarkable acceleration of the catalytic reaction upon near-infrared irradiation is demonstrated, which reveals for the first time the importance of the synergistic effect of photothermal conversion and complex inner structure to the kinetics of the catalytic reduction. The ease of synthesis and fresh insights into catalysis will promote a new platform for novel nanoreactor studies.

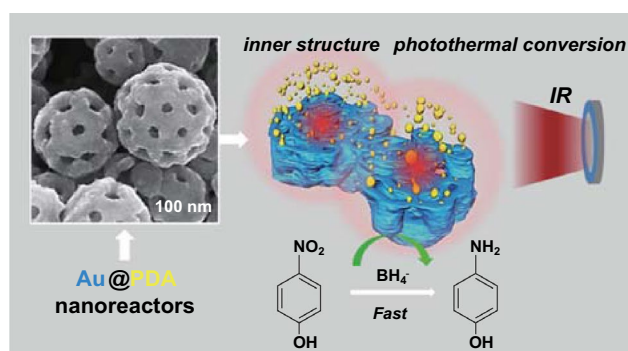

**KEYWORDS** Gold@polydopamine; 3D tomography; Nanoreactor; Catalysis; Photothermal conversion

# 1 Introduction

Photothermal conversion (PTC) nanostructures, which can absorb incident photons and generate heat, have attracted tremendous attention due to their promising applications for photothermal steam generation, photothermal therapy, photothermal imaging, solar energy harvesting, sea water desalinization, and catalysis [1–6]. To date, there has been long-standing interest in developing solar-driven photothermal catalysts to achieve smart and green chemistry, since photothermal conversion is an important pathway to harvest solar energy from the UV to the visible, and even to the infrared (IR) region. Intensive studies have been conducted in the synthesis of PTC materials with various solar absorber materials, e.g., plasmonic metals [7–9], narrow bandgap semiconductors [10, 11], and carbon-based nanostructures [12, 13], which show efficient photothermal conversion. For example, reduced graphene oxide (rGO)-based photothermal catalysts, Ni/rGO and Cu/rGO nanocomposites were reported for the reduction of 4-nitrophenol to 4-aminophenol with $NaBH_4$ as a reducing agent [14, 15]. Under NIR irradiation, the reduction rate was significantly enhanced due to the excellent NIR photothermal conversion property of rGO, which induced heating of the local environment near the Ni or Cu nanoparticles and the surrounding reaction medium. Therefore, the utmost utilization of the heat from photothermal conversion is of great importance to pursue advanced photothermal catalyst.

Apart from photothermal conversion efficiency, unique chemical property of the pristine PTC material together with fine physical structure is an important function that determines the practical application of nanoreactors. Among countless nanostructures created through diverse approaches, compartmentalization is a process abundantly found in nature in order to confine, protect and regulate biological processes and to enable transport of cargo [16–18]. Over the past years, extensive research has been performed to create synthetic analogues of the mostly lipid-based natural compartments, for a variety of applications [19–21]. Fatty acids and phospholipids have been the most commonly used material to construct multi-compartment structures [16–21]. Despite their ability to successfully form vesicles, fatty acids and phospholipids often lack sufficient robustness due to their small bilayer thickness, which leads to low mechanical stability and leakiness. This has triggered research in the area of polymer-based multi-compartment nanostructures.

In principle, such polymer-based photothermal nanoreactors require catalytically active particles to be encapsulated within a permeable shell to separate the inner compartment from the outer bulk solution, and to allow for diffusion of substrates between the outside and the inside of the nanoreactors [22, 23]. Moreover, the polymeric material should enable the ease of constructing into complex nanostructures. Most importantly, the polymer should possess remarkable photothermal property in order to effectively regulate the catalytic behaviors. With the potential to fulfill all of the above-mentioned requirements, PDA maintains its uniqueness over the other types of PTC materials in possessing high conversion efficiency and being a nature-inspired coating. The photothermal conversion efficiency of PDA nanomaterials has been demonstrated to be up to 40% depending on the nanostructure in the near-infrared (NIR) region [24–26]. However, previous studies mainly focus on PDA dense particles and core–shell particles [25, 26]. Few efforts have been made on morphology evolution of PDA particles with higher complexity due to the quick oxidative self-polymerization of dopamine monomer in weak alkaline conditions (pH ~ 8.5) [25]. Compared with other photothermal materials, the relationship between the PDA nanostructures and their photothermal properties, as well as the influence of the complex structure on the kinetics of catalytic reactions under NIR irradiation, has not been intensively studied due to the synthetic difficulties.

Formation of complex features mostly relies on the application of hard templates [27, 28]. But the removal of hard templates usually requires harsh conditions, which may damage the structural integrity. By contrast, soft templates such as oil droplets and polymer micelles are easier to be removed by simple extraction or evaporation [27–29]. Nevertheless, these soft templates only give rise to simple structural features such as hollow particles. Other synthetic techniques such as self-assembly of block copolymers or biomacromolecules largely rely on well-designed supramolecular interactions, which is usually too weak to preserve the shape integrity for sustainable use as nanoreactors [30, 31]. As a result, the pursuit of catalytically active, PTC nanoreactors with multi-compartment structure has been significantly hampered to date.

Herein, we designed the PTC Au@PDA nanoreactors with multi-compartment structure by using a block copolymer (BCP) as template. Au nanoparticles are well known to be catalytically active for reduction reactions [32]. The facile preparation through colloidal routes with well-controlled particle sizes facilities the common use of Au nanoparticles in catalytic reactions. Direct comparison of the kinetics of model reduction reactions among different carrier systems has been available by using Au nanoparticles with similar shape and size [32, 33]. BCP of polystyrene-*b*-poly(2-vinylpyridine) (PS-*b*-P2VP) stands out as a compelling template owing to its tunable microphase-separation structure [34]. Most of all, the ease of removal of the template via direct dissolution in good solvent enables the preservation of the structural integrity. Scheme 1 illustrates the synthesis procedure. By using PS-*b*-P2VP porous particles (obtained from a selective swelling process) as the template, Au nanoparticles are assembled onto the BCP framework through in situ reduction of gold ions under UV irradiation. Under alkaline condition (Tris buffer, pH = 8.5), a layer of polydopamine with controlled thickness can be deposited on the surface of the Au@PS-*b*-P2VP particles. The PS-*b*-P2VP template can be removed by THF, leading to a multi-compartment structure. Electron tomography (ET) has been applied to reveal the 3D structure of the complex particles. The obtained Au@PDA particles have been used as photothermal responsive nanoreactors for the catalytic reduction of 4-nitrophenol. The full kinetics at room temperature has been investigated and compared with other reported catalytic systems in detail. Furthermore, the photothermal effect and catalytic activity of the particles under NIR irradiation have been studied. To the best of our knowledge, this is the first report on the synthesis of PDA-based nanoreactors with multi-compartment structure. These nanoreactors not only exhibit photothermal responsiveness upon NIR, but also provide multi-confinement at nanoscale for target molecules, which may contribute to potential applications in templated synthesis, biotechnology and controllable catalysis.

## 2 Experimental

### 2.1 Commercial Chemicals and Materials

Poly (styrene-2-vinylpyridine) (PS-*b*-P2VP) ($M_n$ (PS) = 50,000 g mol$^{-1}$; $M_n$ (P2VP) = 16,500 g mol$^{-1}$; $M_w/M_n$ = 1.06) was obtained from Polymer Source Inc. Sodium dodecyl sulfate (SDS), dopamine hydrochloride, Gold(III) chloride trihydrate (HAuCl$_4$·3H$_2$O), analytical grade toluene, ethanol, 4-nitrophenol (Nip), and sodium borohydride (NaBH$_4$) were purchased from Sigma-Aldrich and used as received. De-ionized water was used for all aqueous solutions. Tetrahydrofuran (THF) was purchased from Fisher Scientific UK and used as received.

### 2.2 Synthesis of Porous PS-*b*-P2VP Templates

The porous PS-*b*-P2VP nanoparticles have been produced following the reported method [35]. In a typical experiment, 10 mg of PS-*b*-P2VP nanospheres was dissolved into 10 mL toluene and then emulsified into 100 mL water solution with 0.1 wt% of a surfactant (SDS) under sonication. Toluene was evaporated at 75 °C in a water bath, resulting in the solidified PS-*b*-P2VP nanospheres dispersed in the water. The obtained nanoparticles were dispersed in 60 mL ethanol and kept at 75 °C for 60 min to induce the porous structure.

### 2.3 Assembly of Au Nanoparticles onto the Porous BCP Particles and PDA Coating

The porous PS-*b*-P2VP particles (10 mg) were dispersed in 20 mL HAuCl$_4$/ethanol solution (the concentration of HAuCl$_4$ was 10$^{-3}$ M) at room temperature for 12 h to allow

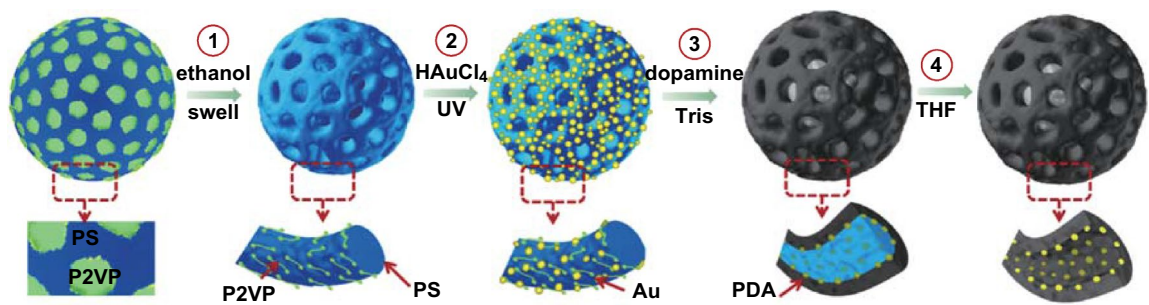

**Scheme 1** Synthesis procedure of the Au@PDA nanoreactors. (1) Selective swelling of PS-*b*-P2VP particles into porous particles; (2) *in situ* growth of gold nanoparticles on the porous PS-*b*-P2VP templates through UV irradiation; (3) deposition of PDA layer onto the porous templates; (4) removal of the PS-*b*-P2VP templates by dissolving in THF

the complexation of the $AuCl_4^-$ ions with the protonated pyridine rings of P2VP. In order to avoid the formation of excess gold nanoparticles in the solution, the complex was separated from the free $AuCl_4^-$ ions by centrifugation. The adsorbed gold ions were reduced into gold nanoparticles by UV light irradiation. Then the obtained Au@PS-*b*-P2VP porous particles were dipped into a solution of dopamine hydrochloride (0.3 mg mL$^{-1}$) dissolved in 10 mL Tris buffer (pH 8.5–8.8). The reaction vessel was kept in ice bath for more than 7 h with vigorous stirring. After the deposition, the samples were collected by centrifugation at 6000 rpm for 10 min. In order to remove the extra dopamine and the secondary particles, the centrifugation should be repeated until the supernatant was colorless.

### 2.4 Generation of Au@PDA Particles

The Au@PDA@PS-*b*-P2VP porous particles were redispersed in THF at room temperature under stirring. After 7 days, the PS-*b*-P2VP frameworks were totally dissolved by THF and the Au@PDA hollow particles remained in the suspension. Repeated centrifugation using THF at 6000 rpm was necessary for the complete removal of the dissolved polymer.

### 2.5 Electron Tomography

Nanoparticles suspended in water were dried on a standard TEM grid (copper grid with a thin layer of carbon). Tomographic data were collected on a JEM-2100 (JEOL GmbH, Eching, Germany) operated at 200 kV and equipped with a 4 k × 4 k CMOS digital camera (TVIPS TemCam-F416). Tilt series were acquired using the Serial-EM acquisition software package [36] with a tilt range of ± 60° and a 2° angular increment at a magnification of 30,000×, corresponding to a pixel size of 3.6 Å at the specimen level.

Tilt series were aligned using patch tracking and then reconstructed using weighted back-projection with the IMOD software package [37]. Surface segmentation was performed using Amira (FEI Company, Eindhoven, The Netherlands). Three-dimensional surface renderings and movies were generated with UCSF Chimera [38].

### 2.6 Photothermal Performance

The photothermal conversion performance of the Au@PDA particles was measured with 1 mL aqueous dispersion at various concentrations (0.025–0.5 mg mL$^{-1}$). The dispersions were placed into several quartz cuvettes and then irradiated with an 808-nm NIR laser for more than 10 min, respectively. Temperature variation of the aqueous dispersions was measured by using a thermometer at each time interval.

### 2.7 Kinetic Study of the Catalytic Reduction of 4-Nitrophenol

In a typical run, sodium borohydride solution (0.5 mL, 0.1 M) was added to a 4-nitrophenol solution (4.5 mL, 0.11 mM) in a glass vessel. The solutions were purged with $N_2$ to remove oxygen from the system before mixing. A certain amount of Au@PDA particles (50 μL, 2.5 mg mL$^{-1}$) was added to the mixed solution and shaken gently before putting into the sample chamber. Au@PS-*b*-P2VP particles with the same content of Au nanoparticles were applied for comparison. The kinetic study of the reaction under NIR irradiation was conducted with 1 ml of the aqueous dispersion (0.1 mg mL$^{-1}$), an 808-nm NIR laser (3 W cm$^{-2}$) for more than 10 min.

The kinetic analysis was performed by UV–Vis spectroscopy at 400 nm (PerkinElmer, Lambda 650 spectrometer). The nanoparticles size was measured from TEM image and is shown in Fig. S6a. The content of Au nanoparticles was estimated from TGA results (Fig. S6b). For this calculation, the Au nanoparticles were assumed as spheres and the density of bulk Au metal was used ($\rho = 19.28$ g cm$^{-3}$).

The evaluation of the data was done using a MATLAB sheet as reported by Gu et al. [32]. The concentration of Nip as the function of reaction time, $c_{Nip,exp}$, was analyzed by a numerical solution of a couple of equations by two MATLAB routines as reported by Gu et al. [32]. The MATLAB routines were used to calculate the theoretical Nip concentration $c_{Nip,th}$ as the function of time for a given values of $K_{Nip}$, $K_{BH4}$, $K_{Hx}$, $k_a$, $k_b$, and n. These data are compared to the experimental results, and the constants are changed until agreement with the experiment is reached.

### 2.8 Materials Characterization

The obtained nanostructures were examined by transmission electron microscopy (TEM) using a JEOL JEM-2100 (JEOL GmbH, Eching, Germany). The size distributions of Au nanoparticles were determined from TEM images using the ImageJ software [39]. More than 100

Au nanoparticles were counted. For the measurement of UV–Vis absorption spectra, the reaction solution with Au@PDA particles as catalyst was placed in a quartz sample cell with a 1.0-cm cell path length. UV–Vis spectra (at 400 nm) were recorded by using Lambda 650 spectrometer supplied by PerkinElmer at 20 °C with reference spectrum of the Au@PDA particles in water. TGA measurements were taken using a Netzsch STA409PC LUXX from 25 to 600 °C under a constant argon flow (30 mL min$^{-1}$) with a heating rate of 10 K/min. Nitrogen adsorption experiments were performed with a Quantachrome Autosorb-1 at liquid nitrogen temperature, and data analysis was performed by Quantachrome software. The specific surface area was calculated using the Brunauer–Emmett–Teller (BET) equation. Pore size distribution was determined by Barrett–Joyner–Halenda (BJH) method. Samples were degassed at 50 °C for 24 h before measurements.

## 3 Results and Discussion

### 3.1 Formation of the Au@PDA Particles with Multi-compartment Structure

In this study, PS-*b*-P2VP particles are produced via evaporation of the (PS-*b*-P2VP/toluene)-in-water emulsion. Figure 1a, b presents the porous PS-*b*-P2VP structures obtained from selective swelling of the dense particles by ethanol. Serving as template for complex nanomaterials, the particles are required to possess durable framework and open porous structure, which are dominated by the appropriate molecular weight ratio ($r$) of the swellable minority to the solid majority. Thus, PS-*b*-P2VPs with different $r$ ($M_n$ (P2VP)/$M_n$ (PS)) values are studied. It is found that a perfect porous structure can be obtained at a moderate swelling ratio of ($M_n$ (P2VP) = 16,500 g mol$^{-1}$, $M_n$ (PS) = 50,000 g mol$^{-1}$, $\frac{M_{n(P2VP)}}{M_{n(PS)}} = 0.32$). As shown in Fig. 1a, b nanosphere with an interconnected porous network architecture is obtained, as proven by the TEM analysis of this sample shown in Fig. S1a. These particles not only keep the spherical shape but also exhibit an open porous structure. With a higher swelling ratio ($M_n$ (P2VP) = 10,400 g mol$^{-1}$, $M_n$ (PS) = 23,600 g mol$^{-1}$, $\frac{M_{n(P2VP)}}{M_{n(PS)}} = 0.44$), the resulted porous structure is fragile and easily collapsed (Fig. S1b), while at a relatively lower swelling ratio ($M_n$ (P2VP) = 4800 g mol$^{-1}$, $M_n$ (PS) = 26,000 g mol$^{-1}$, $\frac{M_{n(P2VP)}}{M_{n(PS)}} = 0.18$), a closed multilayers structure formed (Fig. S1c), which will block the functionalization of the inner part. Thus, the PS$_{50000}$–P2VP$_{16500}$ is used as the model polymer to produce the porous templates.

The PS-*b*-P2VP framework contains pyridine units, from which the N atoms with lone pair electrons can serve as electron donors. Various types of metal precursors can be coordinated with them, which can be further reduced to respective metal nanoparticles [40]. Figure 1c shows the full loading of Au nanoparticles onto the porous PS-*b*-P2VP templates by UV reduction of Au ions. Au nanoparticles with a diameter of ~9 nm have been homogenously deposited on the porous templates.

The primary advantage of using PDA is that it can be easily deposited on both inorganic and organic substrates,

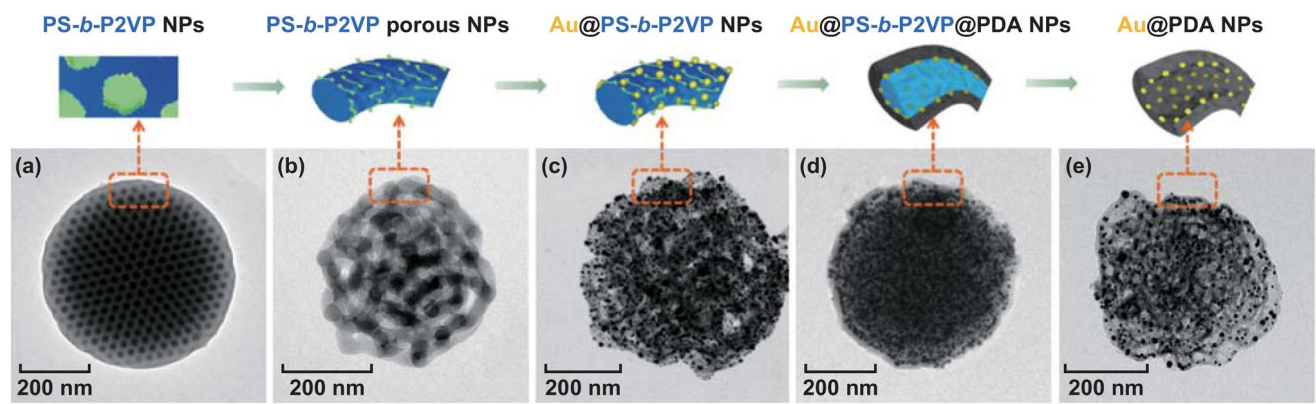

**PS-*b*-P2VP NPs**    **PS-*b*-P2VP porous NPs**    **Au@PS-*b*-P2VP NPs**    **Au@PS-*b*-P2VP@PDA NPs**    **Au@PDA NPs**

(a)  200 nm   (b)  200 nm   (c)  200 nm   (d)  200 nm   (e)  200 nm

**Fig. 1** TEM images of **a** PS-*b*-P2VP particle with ordered microphase-separation structure; the dark dots present the I$_2$-stained P2VP domains, **b** porous PS-*b*-P2VP particles derived from selective swelling, **c** Au@PS-*b*-P2VP particle, **d** Au @PS-*b*-P2VP@PDA particle, and **e** Au@PDA particle with multi-compartment structure after removal of the PS-*b*-P2VP template

including superhydrophobic surfaces. Catechol has been postulated to play an important role in the adhesion [25]. This adherence is essential to trap Au nanoparticles, which is difficult to conduct with other polymers unless additional surface modification of the Au nanoparticles, is introduced. Another advantage is that dopamine can self-polymerize under weakly alkaline condition at room temperature [25, 41]. This mild condition is essential for the fidelity retention of the porous structure of the PS-*b*-P2VP templates, which can be easily destroyed/extracted in organic solvents, acid solution, and high temperature [35].

When dopamine monomers are added into an alkaline solution, the polymerization of dopamine monomers immediately occurs, coupled with a color change from colorless to pale brown, and finally deep brown along the reaction time. Figure 1d shows the PDA-coated Au@ PS-*b*-P2VP particles. We find that the porous structure is no longer obvious. This is because PDA has similar contrast to PS-*b*-P2VP, and the porous space is almost filled by the PDA layer. The

PS-*b*-P2VP template can be removed easily by dissolution in THF at room temperature. Figure 1e shows clearly that the multi-compartment structure of PDA is generated after the removal of PS-*b*-P2VP. The gold nanoparticles originally deposited on the PS-*b*-P2VP framework are fully trapped into the PDA tubes. No aggregation of the gold nanoparticles is found in the TEM image, indicating that the PDA supporting layer can effectively avoid the aggregation of the Au nanoparticles.

Figure 2 shows the comparison of PDA porous spheres without (Fig. 2a–c) and with gold nanoparticles (Fig. 2d–f). The tubular structure of PDA can be clearly seen in Fig. 2c, which is due to the replication of PDA from the PS-*b*-P2VP porous structures. In Fig. 2f, gold nanoparticles with an average diameter of 9.8 nm can be observed in the hollow PDA nanotubes. Owing to the good affinity of PDA to both organic and inorganic materials, the gold nanoparticles are totally embedded inside the PDA layers. This can be demonstrated by the SEM

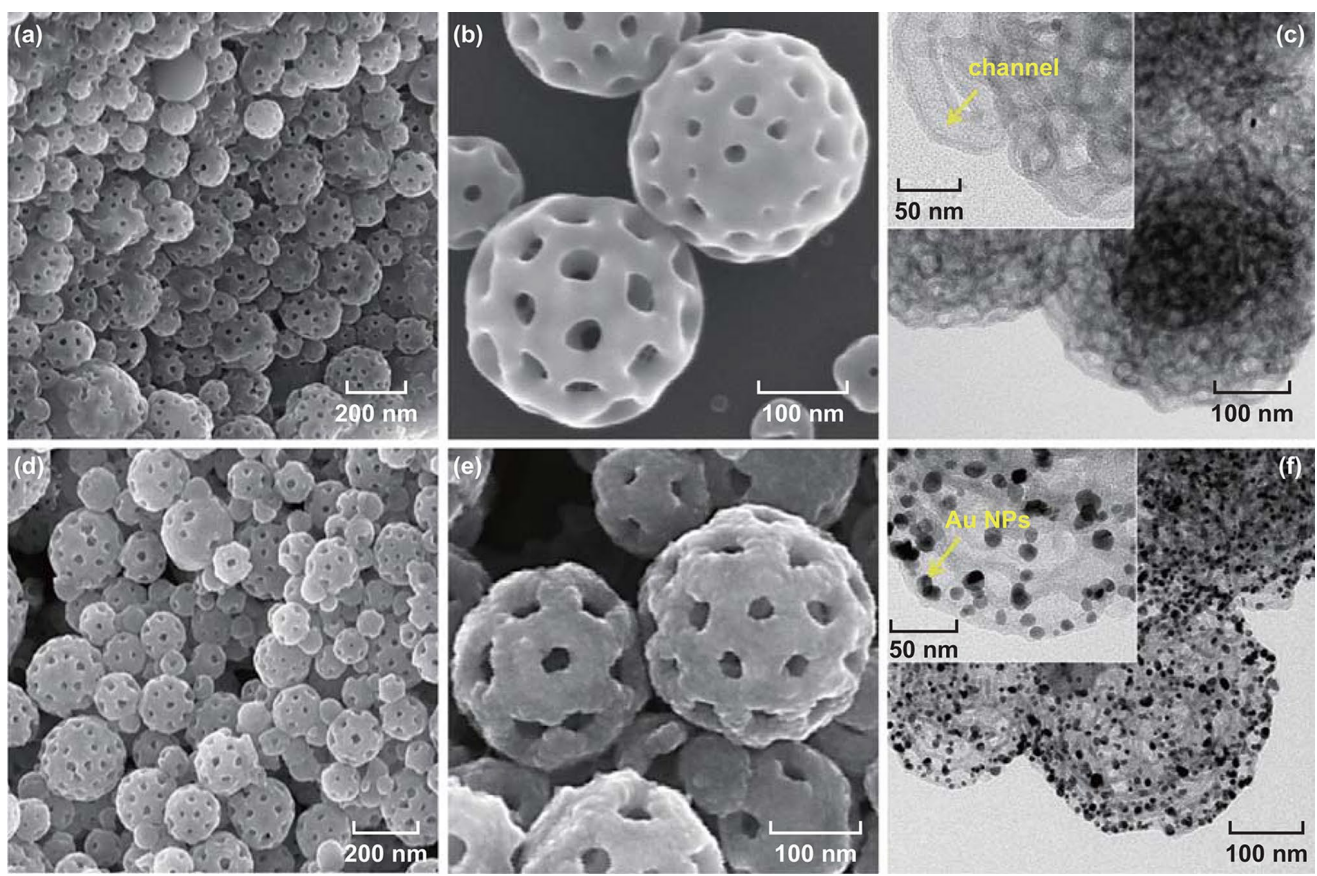

**Fig. 2 a**, **b** SEM and **c** TEM images of the PDA porous particles. **d**, **e** SEM and **f** TEM images of the Au@PDA particles. Insets are the enlarged parts of the particles

  

images (Fig. 2b, e), where higher surface roughness can be observed in Au@PDA compared to the pure PDA particles. This is derived from the surface loading of Au NPs onto PS-*b*-P2VP templates, which is further coated by the PDA layers. Further characterizations by FTIR and XRD demonstrate the formation of PDA and gold nanoparticles, respectively. As shown in Fig. S2a, the absorption bands located at 1490, 1435, and 1250 cm$^{-1}$ indicate the formation of polyphenol structure [42]. The absorption at 1622 cm$^{-1}$ is assigned to the stretching mode of aromatic C–C bonds of indole [43]. Figure S2b shows four diffraction peaks at 38.2°, 43.0°, 64.6°, and 78.1°, which are assigned to the (111), (200), (220), and (311) planes of the face-centered cubic (*fcc*) gold, respectively. The BET specific surface area of the Au@PDA particles is 62.2 m$^2$ g$^{-1}$ (Fig. S2c). The pore size distribution curve shows a narrow range of mesopores between 3 and 5 nm (centered at 3.3 nm) and a broad range between 11 and 24 nm.

The robustness of the Au@PDA scaffolds is especially important when they are applied as reusable nanoreactors, which is mainly dictated by the thickness of the PDA layer. In our work, the PDA thickness can be controlled effectively by polymerization time. Different reaction times have been applied for the PDA coating process by keeping the dopamine concentration constant (0.3 mg mL$^{-1}$). When the reaction time is less than 4 h, a very thin layer of PDA (2–3 nm) can be observed covering on the surface of the PS-*b*-P2VP particles (Fig. S3a) before removal of the template. But the PDA porous particles produced here are fragile and significantly distorted upon centrifugation and sonication (Fig. S3b). Longer reaction time to 11 h results in thicker PDA layers of about 14 nm (Fig. S4). Thus, we increase the reaction time gradually from 5 to 11 h to generate durable Au@PDA nanoreactors with tunable PDA thickness. As is shown in Fig. 3, increasing the polymerization time from 5, 7, to 11 h leads to the increase

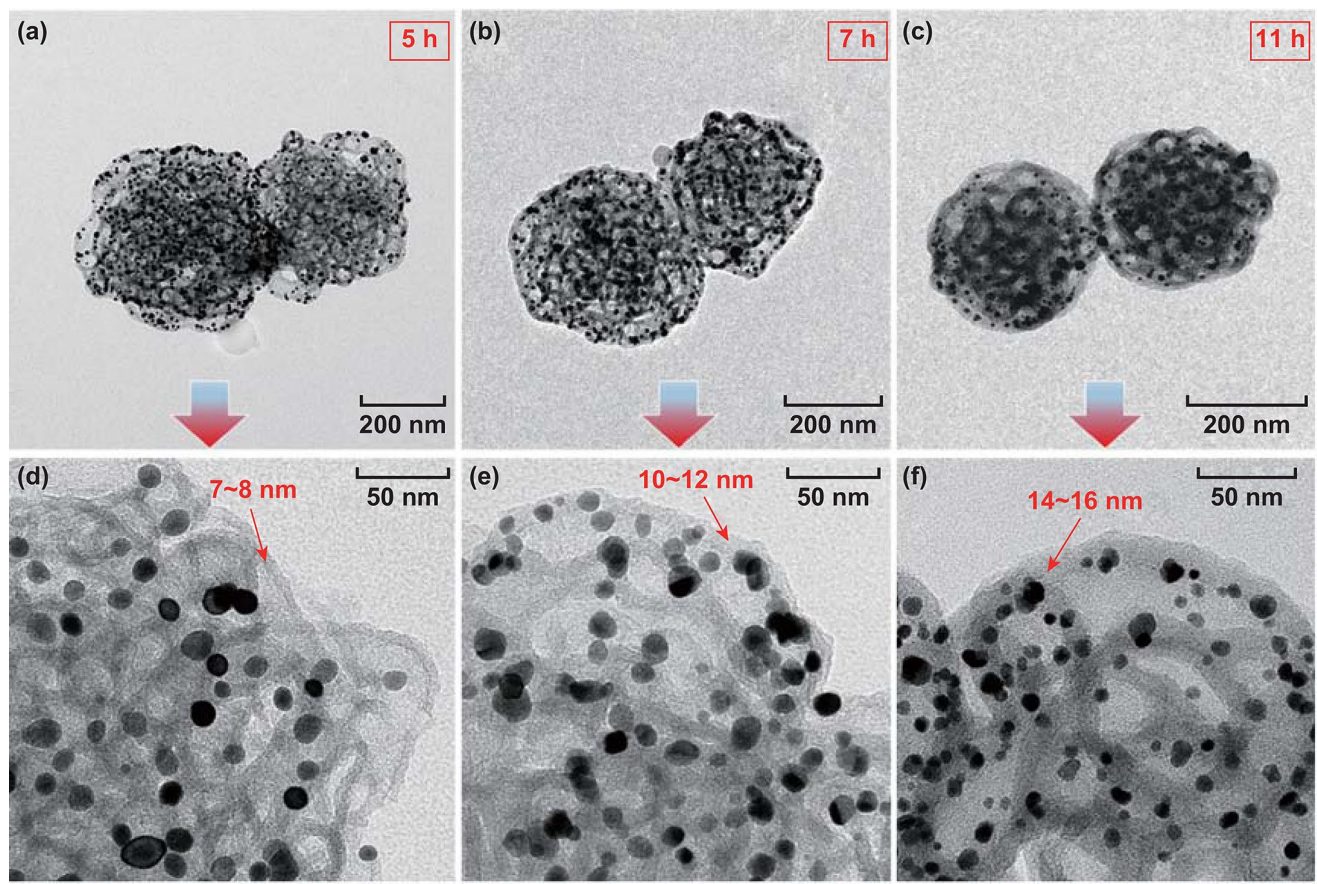

**Fig. 3** TEM images of the Au@PDA porous particles with different thicknesses of the PDA layer by adjusting the reaction time from 5 h, 7–8 nm (**a, d**) to 7 h, 10–12 nm (**b, e**), and 11 h, 14–16 nm (**c, f**)

in the PDA thicknesses from 7–8, 10–12, to 14–16 nm. The uniform PDA layers can be clearly seen from TEM images with high magnification (Fig. 3d–f), indicating the structural integrity of the porous particles.

### 3.2 3D Visualization of the PS-*b*-P2VP Template and the Au@PDA Particle

For materials with complex 3D nanostructures, micrographs obtained by TEM are often inconclusive as the structural features are superimposed along the electron beam direction. Electron tomography (ET) offers the possibility to overcome that limitation by obtaining a reconstruction of the object's 3D structure from a series of two-dimensional micrographs [44, 45]. In our study, we successfully utilized ET for the detailed structural analysis of both the porous PS-*b*-P2VP templates and the Au@PDA particles. A normal TEM image of the PS-*b*-P2VP template shown in Fig. 4a indicates the porous structure of the particles. Figure 4b is a gallery of selected XY slices through the tomographic reconstruction of the particle in Fig. 4a. 3D surface renderings of the particle present both the surface and the internal structure of the PS-*b*-P2VP template (Fig. 4c–e), where PS-*b*-P2VP micellar fibers (27 nm in width, measured in Fig. 4b) form an interconnected network. Such highly porous structure provides large surface for the deposition of PDA, which leads to multi-compartment structure after removal of the template (Fig. 5).

Figure 5a, b shows the normal TEM images and selected XY slices through the tomographic reconstruction of the Au@PDA particles, respectively. The tube in Fig. 5b is measured as 27 nm in width, which is consistent with that of the PS-*b*-P2VP micellar fibers (Fig. 4b). The tubes are confirmed to be continuous inside the particles, which is attributed to the precise replication of the PS-*b*-P2VP micellar fibers by PDA and the subsequent removal of the template. Figure 5c shows the branches of the PDA tubes; no gold nanoparticles can be observed from the surface view, while the cutting plane views (Fig. 5d1–d4) and the semitransparent views (Fig. 5e–g) show that gold nanoparticles are distributed all over the inner space of the particles, signifying that gold nanoparticles are fully trapped in the hollow PDA tubes and distributed homogeneously without detectable aggregation. Furthermore, Fig. 5d2, d4 elucidates the spatial

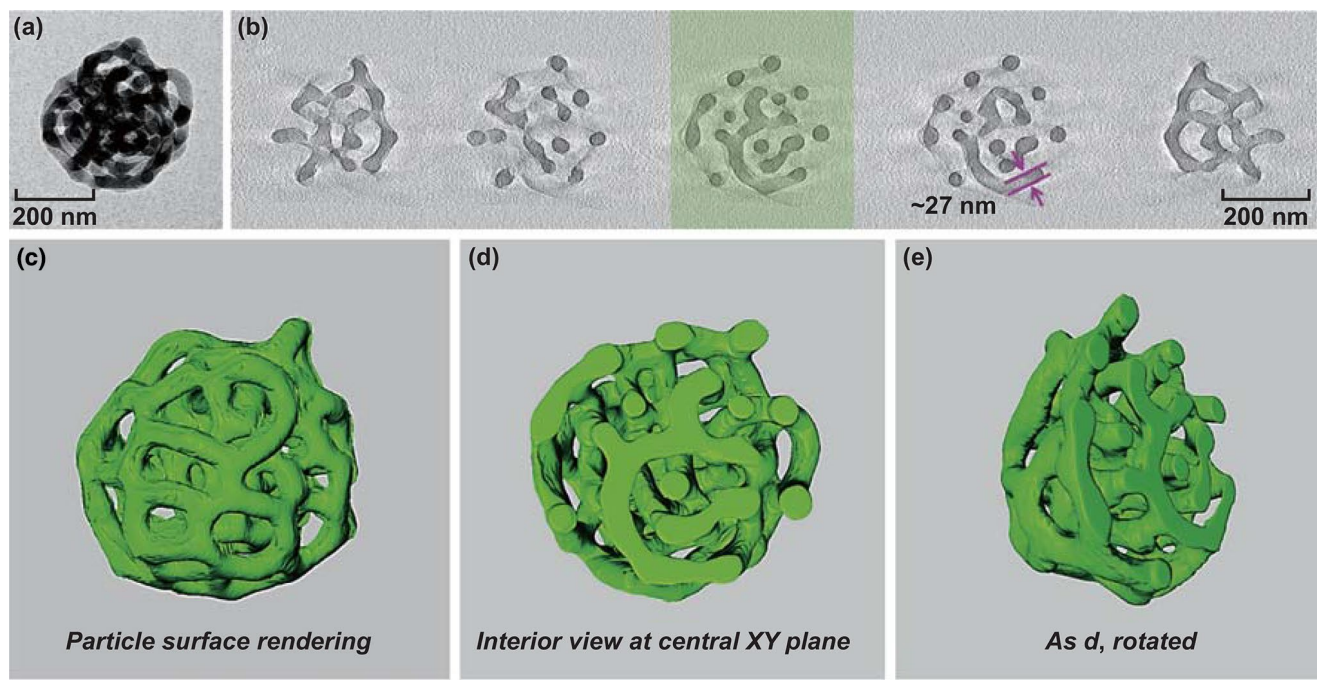

**Fig. 4** TEM analysis of the PS-*b*-P2VP templates: **a** normal TEM image and **b** selected XY slices through the tomographic reconstruction of an individual PS-*b*-P2VP porous particle. The highlighted green slice in **b** corresponds to the central slice **d**. **c–e** 3D particle surface renderings with and without interior views, using a cutting plane through the central XY slice (PS-*b*-P2VP shown in green). (Color figure online)

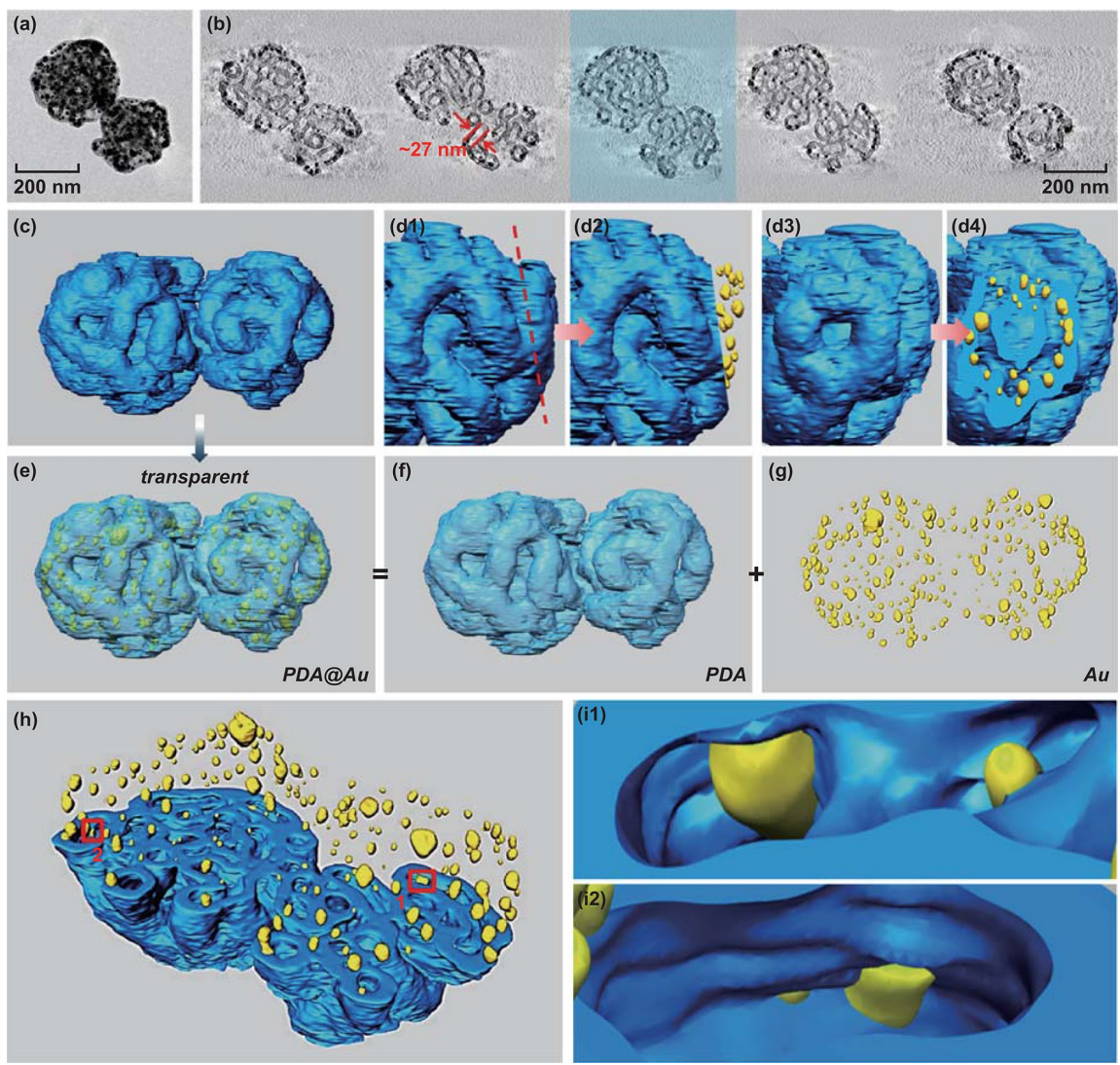

**Fig. 5** TEM analysis of the Au@PDA particles: **a** normal TEM image and **b** selected XY slices through the tomographic reconstruction of two adjacent Au@PDA particles. **c–i** 3D surface renderings of the particles (PDA shown in blue and Au shown in yellow): **c** surface view (**d1**, **d2**) subregion including cut away view to highlight the interior distribution of gold nanoparticles in PDA (**d3**, **d4**) rotated view of d1, d2, respectively. **e** as **c** but with semitransparent PDA layer to show internal structure. **f**, **g** as **e** but components of PDA and Au are shown separately. **h** as **c** but with a cutting plane through the central XY slice of PDA. **i1**, **i2** magnified views of selected marked regions in **h**. (Color figure online)

distribution of the gold nanoparticles along the PDA tubes. Enlarged views of the selected regions (Fig. 5i1, i2) from Fig. 5h further prove the encapsulation of gold nanoparticles inside the PDA tubes. All these results demonstrate that ET is powerful to give a direct proof of the multi-compartment structure of the Au@PDA particles. A video of the reconstruction can be found in the supplementary materials.

### 3.3 Kinetic Study of the Reduction of 4-Nitrophenol

Owing to the permeable property of PDA thin layers, the Au@PDA particles can be used as nanoreactors for the full kinetic study of reduction reactions. The catalytic reduction of 4-nitrophenol (Nip) by $NaBH_4$ to 4-aminophenol at 20 °C has been used as a model system, which can be precisely followed by the UV–Vis spectroscopy as shown in Fig. S5a [46]. The characteristic absorption at 400 nm reduces with time due to the conversion of 4-nitrophenol. Another characteristic absorption at about 290 nm appears and increases with time due to the generation of the end product, 4-aminophenol. The Au@PDA particles with PDA thickness of 8 nm have been applied as catalytic nanoreactors (Figs. S5 and S6). As shown in Fig. S5b, the nanoreactors present good reusability. After 5 repeated cycles, they show almost identical activities without obvious decrease in the conversion. Comparing the TEM images of the Au@PDA particles before and after the reaction, no obvious change in size and size distribution of the Au particles has been found (Fig. S5c, d).

Analysis of the total reaction rate shows that the reaction is surface-controlled in this system (see the theory part of the Supplementary materials). The reason is that the total catalytic time is much slower than the diffusive approach of the reactants to the dense layer of nanoparticles at the nanoreactor surface. The evaluation of the kinetic data was done using a MATLAB sheet as reported by Gu et al. [32, 33, 47]. The kinetic study follows a system of two coupled differential equations, which describe the two steps of the reduction (Fig. S7):

$$-\frac{dc_{Nip}}{dt} = k_a S \frac{\left(K_{Nip}\right)^n \left(c_{Nip}\right)^n K_{BH4} c_{BH4}}{\left[1 + K_{Hx} c_{Hx} + \left(K_{Nip} c_{Nip}\right)^n + K_{BH4} c_{BH4}\right]^2}$$
$$= \left(\frac{dc_{Hx}}{dt}\right)_{source} \tag{1}$$

Here, $K_{Nip}$, $K_{Hx}$, and $K_{BH4}$ are the Langmuir adsorption constants of the respective compounds and $k_a$ and $k_b$ represent the reaction rate constants of step A (4-nitrophenol is first reduced to 4-nitrosophenol and then to 4-hydroxylaminophenol (Hx)) and step B (Hx is reduced to Amp), respectively. Excellent agreement between theory and experiment is found (Figs. 6a, b and S8). The thermodynamic and kinetic parameters including $K_{Nip}$, $K_{Hx}$, $K_{BH4}$, and $k_a$, $k_b$ have been obtained from the fitting results (Fig. 6c, d) and compared with reported systems (Table 1).

For the rate constants $k_a$ and $k_b$, the values obtained in this study are smaller than that of the compared systems. The decrease in $k_a$ and $k_b$ may be related to the confinement of the PDA tubes, inside of which Au nanoparticles are encapsulated and part of their surface is occupied. It is also known theoretically that the partitioning of reactants inside the nanoreactor scales the surface reaction rate constant and the excluded volume of the PDA may lead to partition ratios smaller than unity [48]. In addition, the Au@PDA nanoreactors have a smaller $K_{Nip}$, similar $K_{BH4}$ and $K_{Hx}$. The slightly smaller $K_{Nip}$ indicates that the Au@PDA particles are not as favorable as that of other reported systems for the adsorption of reactants. However, $K_{BH4}$ and $K_{Hx}$ of the Au@PDA nanoreactors are quite similar to those of Au and Pd nanoparticles immobilized in the spherical polyelectrolyte brushes (SPB), and ligand-free Au nanoparticles. The difference is within the experimental error. This indicates that the complex structure of the PDA particles has no obvious effect on the adsorption of $BH_4^-$ and Hx. Hence, we conclude that the Au@PDA nanoreactors show significant influence on $k_a$, $k_b$ and $K_{Nip}$, but little effect on the adsorption of $BH_4^-$ and Hx. Thus, the Au@PDA particles are proved to be catalytically active nanoreactors for the reduction of 4-nitrophenol confined in nanotubes, revealing different kinetics from the reported colloidal catalyst systems.

### 3.4 Enhanced Catalytic Behavior of the Au@PDA Nanoreactors Under NIR Irradiation

Prior to investigating the influence of the photothermal effect on the reaction, the photothermal conversion efficiency ($\eta$) of the pure PDA porous particles was measured to study the photothermal property of the complex structures (Fig. 7a–c). The $\eta$ value was calculated according to the previously

$$\frac{dc_{Hx}}{dt} = k_a S \frac{\left(K_{Nip}\right)^n \left(c_{Nip}\right)^n K_{BH4} c_{BH4}}{\left[1 + K_{Hx} c_{Hx} + \left(K_{Nip} c_{Nip}\right)^n + K_{BH4} c_{BH4}\right]^2}$$
$$- k_b S \frac{K_{Hx} c_{Hx} K_{BH4} c_{BH4}}{\left[1 + K_{Hx} c_{Hx} + \left(K_{Nip} c_{Nip}\right)^n + K_{BH4} c_{BH4}\right]^2}$$

reported method [26] as Eq. 3:

$$\eta = \frac{hA\Delta T_{max} - Q_s}{I\left(1 - 10^{-A_\lambda}\right)} \tag{3}$$

where $h$ is the heat transfer coefficient, $A$ is the surface area of the container, $\Delta T_{max}$ is the temperature change of the PDA porous particles solution at the maximum steady-state temperature, $I$ is the laser power, $A_\lambda$ is the absorbance of the PDA porous particles solution at 808 nm, and $Q_s$ is the heat associated with the light absorbance of the solvent. $hA$ is

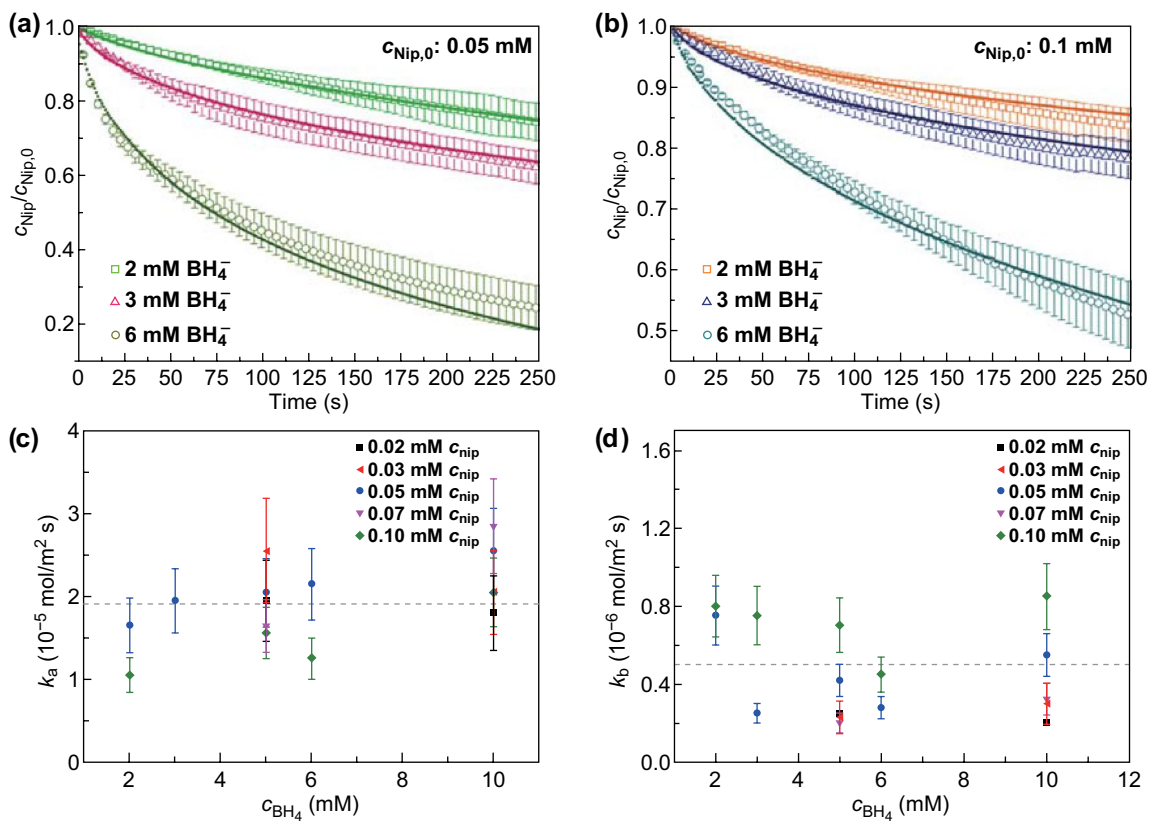

**Fig. 6** **a, b** Fit of the concentration of 4-nitrophenol as the function of time by the theoretical model. The concentration of 4-nitrophenol was normalized to the respective starting concentration $c_{Nip,0}$. The experimental data have been taken from reactions at 20 °C (data points with error bars). The solid lines refer to the fits by the kinetic model. **c, d** Kinetic constants $k_a$ and $k_b$ derived from fitting results. The dash lines indicate the average values

**Table 1** Summary of the kinetic and thermodynamic parameters of the kinetic analysis

| Parameter | $k_a$ (mol/m² s) | $k_b$ (mol/m² s) | $K_{Nip}$ (L/mol) | $K_{BH4}$ (L/mol) | $K_{Hx}$ (L/mol) | $n$ (Nip) | $d_{NPs}$ (nm) |
|---|---|---|---|---|---|---|---|
| This study | $1.9 \pm 0.6 \times 10^{-5}$ | $0.5 \pm 0.3 \times 10^{-6}$ | $1200 \pm 400$ | $80 \pm 20$ | $(190 \pm 20) \times 10^3$ | 0.5 | 9.8 |
| SPB-Au[32] | $9.4 \pm 2.6 \times 10^{-4}$ | $5.6 \pm 1.4 \times 10^{-5}$ | $3700 \pm 900$ | $50 \pm 4$ | $(160 \pm 15) \times 10^3$ | 0.5 | 2.6 |
| SPB-Pd[47] | $1.4 \pm 0.4 \times 10^{-4}$ | $0.8 \pm 0.6 \times 10^{-5}$ | $2000 \pm 600$ | $70 \pm 10$ | $(180 \pm 20) \times 10^3$ | 0.5 | 2.5 |
| Au[33]$_{Ligand-free}$ | $5.8 \pm 3.1 \times 10^{-4}$ | $5.4 \pm 2.0 \times 10^{-5}$ | $1800 \pm 700$ | $60 \pm 10$ | $(160 \pm 25) \times 10^3$ | 0.5 | 7.2 |

*SPB* spherical polyelectrolyte brushes

determined by applying the linear time data from the cooling period versus $-\text{Ln}\theta$ (Fig. 7c, where $\theta$ is defined as $\frac{\Delta T}{\Delta T \max}$ according to the reported calculation approach) [26]. Therefore, the $\eta$ value of pure PDA porous particles is determined to be 19.5%, which is comparable among various PTC nanomaterials, revealing the significant impact of the porous structure on the photothermal conversion in aqueous system.

The photothermal property of the Au@PDA particles has been further studied systematically to validate their potential as photothermal nanoreactors. Different particle concentrations and NIR laser powers have been taken into account to evaluate the photothermal conversion. As shown in Fig. 7d, e, the amount of generated heat depends on both the particle concentration and the light power. When the irradiation power is kept constant with an 808-nm laser at 3 W cm⁻² for

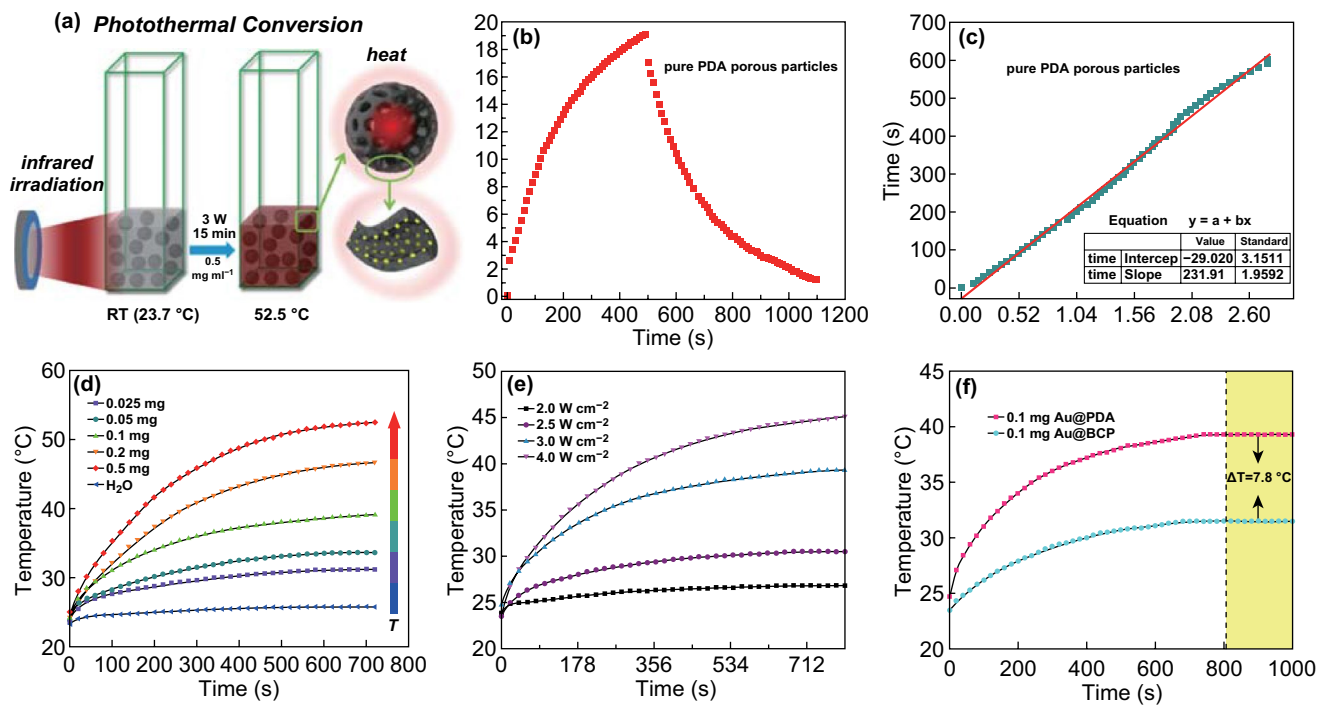

**Fig. 7** **a** Schematic diagram of photothermal conversion of the Au@PDA particles. **b** The photothermal response of the PDA porous particles in aqueous solution (0.2 mg mL$^{-1}$) for 500 s with a NIR laser (808 nm, 2 W cm$^{-2}$), and then, the laser was shut off. **c** Linear time data versus $-\ln\theta$ obtained from the cooling period of **b**. **d** Temperature changes of pure water and various concentrations of Au@PDA particles dispersions under irradiation with NIR light (808 nm, 3 W cm$^{-2}$). **e** Temperature dependence on the irradiation power with constant particle concentration (0.1 mg mL$^{-1}$). **f** Comparison of the photothermal effects of Au@PDA and Au@P2VP particles with 0.1 mg mL$^{-1}$ at 3 W cm$^{-2}$

12 min, the temperature of the aqueous Au@PDA dispersion increases from 23.7 to 31.2, 33.6, 39.3, 46.7, and 52.5 °C with the increase in particle concentration from 0.025 to 0.05, 0.1, 0.2, and 0.5 mg mL$^{-1}$, respectively (Fig. 7d). As a comparison, pure water is measured under the same irradiation condition, which only shows tiny change in temperature from 23.7 to 25.8 °C. When the particle concentration is kept constant at 0.1 mg mL$^{-1}$, the temperature increases from 23.7 to 26.8, 30.5, 39.3, and 45.1 °C with increased irradiation power from 2 to 2.5, 3, and 4 W cm$^{-2}$, respectively (Fig. 7e). In addition, Au nanoparticles have been also reported for their photothermal properties. In order to figure out the photothermal contribution of the Au nanoparticles in the composite particles, Au@PS-b-P2VP particles have been measured and compared with the Au@PDA particles at the same concentration (0.1 mg mL$^{-1}$) and irradiation power (3 W cm$^{-2}$). Figure 7f shows a maximum temperature of 31.5 °C after irradiation for 15 min in the case of Au@PS-b-P2VP particles, which is much lower than that of Au@PDA particles (39.3 °C).

To study the catalytic behavior of the Au@PDA nanoreactors under NIR irradiation, three catalytic reactions with the same amount of catalysts (0.1 mg mL$^{-1}$) have been conducted: (1) reaction without NIR irradiation at room temperature (RT); (2) reaction under NIR irradiation at 3 W cm$^{-2}$ for 15 min, which leads to an increase in the solution temperature to 39.3 °C; and (3) reaction without NIR irradiation at 39.3 °C (heated with water bath). Figure 8a–c shows the catalytic activities of the three systems. Figure 8d shows the multi-compartment structure of the nanoreactors after being used for catalytic reactions under NIR irradiation, suggesting the retention of the structure integrity. Since the reaction under NIR irradiation is extremely fast, the conversions of different reactions have been only calculated and compared for the first 40 s, which are 1.1% (without NIR irradiation at RT), 48.4% (without NIR irradiation at 39.3 °C) and 89.9% (under NIR irradiation), respectively. We found that the reactions without NIR irradiation at 39.3 °C and that under NIR irradiation are much faster than the reaction at RT (Fig. 8a, the UV–Vis spectra at 0 and 40 s almost overlap with each other). More interestingly, although the two reactions in Fig. 8b proceed at the same bulk temperature (39.3 °C), the one under NIR irradiation (Fig. 8b, red dots) is much faster than the reaction heated with water bath without NIR irradiation. For the sake of clarity, control experiment with only reactants of 4-nitrophenol and BH$_4^-$ but without Au@PDA nanoreactors under the same NIR power and irradiation time has been conducted. Figure S10 shows that 4-nitrophenol is stable under NIR irradiation; no degradation is found in the UV–Vis spectra from 0 to 40 s and even 500 s. This paves the way to attribute the significant acceleration of the reaction to the nanoreactors. On the basis of the same bulk temperature (39.3 °C), we speculate that the local surface temperature of the Au nanoparticles under NIR irradiation is higher than that heated with water bath. When the bulk solution is heated up to 39.3 °C, the local temperature of the PDA nanoreactors should be much higher since PDA works efficiently as a heat source. The preservation of the local heat on the surface of Au nanoparticles by the PDA layers may be crucial to generate a continuous higher temperature in the nanoreactor than the bulk solution. This is similar with organisms in nature [49] that a heterogeneous thermal environment can be created in the reaction under NIR irradiation, where a deviation of local particle temperature from bulk solution temperature exists over space and time, stemming from radiative and conductive exchanges of heat. In addition, the multi-compartment structure may further preserve the heat generated from photothermal conversion of PDA. Au nanoparticles are likely protected inside a heating compartment with multiple heating layers, where a higher local temperature can be sustained. Moreover, the Au nanoparticles can work as photothermal converting agent as well, which is probably beneficial to a higher surface temperature. Thus, the synergistic effect of NIR-induced photothermal conversion as well as the complex inner structure of the Au@PDA nanoreactors account for the faster kinetics, to the best of our knowledge, which is revealed for the first time. Given that the PDA complex structure acts a pivotal part in enhancing the catalytic activity under NIR irradiation, the relationship between the structural features and the photothermal properties, such as the photothermal conversion efficiency and the heat preservation effect, has become of great importance to be revealed, which will be our focus in a future work with systematically experimental and theoretical studies.

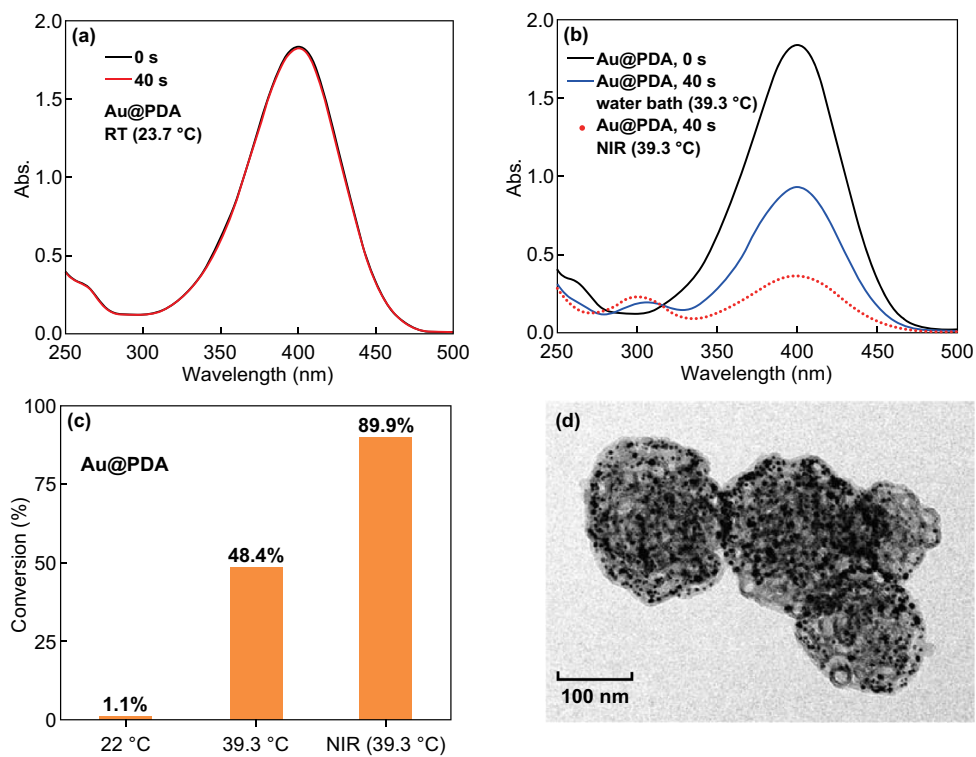

**Fig. 8** UV -Vis spectra of the reduction of 4-nitrophenol using Au@PDA particles as nanoreactors **a** without NIR irradiation at RT, **b** without NIR irradiation at 39.3 °C and under NIR irradiation at 3 W cm$^{-2}$ for 15 min. **c** Conversions of the three reactions. **d** TEM image of Au@PDA particles after used in catalytic reactions under NIR irradiation. (Color figure online)

## 4 Conclusion

We present for the first time the synthesis of PTC Au@PDA nanoreactors with multi-compartment structure by using PS-*b*-P2VP porous particles as soft template. The preferential complexation of P2VP blocks with gold ions facilitates the deposition of gold nanoparticles. The good affinity of PDA to universal materials allows its accurate replication of the PS-*b*-P2VP template with a complex mesoporous structure. Furthermore, the permeable property of the PDA layer is essential for the removal of the template by simple dissolution in THF, which contributes to the preservation of both the chemical property and the integral complex structure. 3D reconstruction of the particles confirms the multi-compartment structure, where Au nanoparticles are encapsulated evenly inside the particles. The composite particles can be applied as model nanoreactors for the catalytic reduction of 4-nitrophenol. The kinetic results show different $k_a$, $k_b$, and $K_{Nip}$, but similar $K_{BH4}$ and $K_{Hx}$ in comparison with Au nanoparticles immobilized in SPB and ligand-free Au nanoparticles. Notably, for the first time, we studied the importance of the synergistic effect of photothermal conversion and complex inner structure on the catalytic reduction. A higher surface temperature of the Au nanoparticles under NIR irradiation maybe generated and preserved by multiple PDA layers, which enables a remarkable acceleration of the reduction reaction. The unique photothermal property of PDA and its complex inner structure underlines the potential as a new class of catalytic nanoreactors.

**Acknowledgements** We acknowledge support from the DFG through SFB 951 Hybrid Inorganic/Organic Systems for Opto-Electronics (HIOS). JD acknowledges funding by the European Research Council (ERC) Consolidator Grant with Project Number 646659-NANOREACTOR. The authors also thank the Joint Lab for Structural Research at the Integrative Research Institute for the Sciences (IRIS Adlershof).